\documentclass[sigconf]{acmart}
\usepackage{tikz}
\usepackage{caption}
\AtBeginDocument{%
  }

\setcopyright{acmlicensed}
\copyrightyear{2026}
\acmYear{2026}
\acmDOI{XXXXXXX.XXXXXXX}
\acmConference[KDD 2026]{The 32nd ACM SIGKDD Conference on Knowledge Discovery and Data Mining}{June 03--05, 2026}{Jeju, Korea}
\acmISBN{978-1-4503-XXXX-X/2026/06}

\begin{document}

\title{Beyond Interleaving: Causal Attention Reformulations for Generative Recommender Systems}

\author{Hailing Cheng}
\email{haicheng@linkedin.com}
\orcid{1234-5678-9012}
\affiliation{%
  \institution{Linkedin Inc}
  \city{Mountain View}
  \state{California}
  \country{USA}
}

\begin{abstract}
Generative recommender systems (GR), exemplified by Meta’s HSTU ranker architecture, model user behavior as a sequence generation problem by interleaving item and action tokens. While effective, this formulation introduces fundamental limitations: it doubles sequence length, incurs quadratic computational overhead, and relies on implicit attention mechanisms to recover the causal relationship that an item interaction $i_n$ elicits a user action $a_n$. Moreover, interleaving heterogeneous item and action tokens forces Transformers to disentangle semantically incompatible signals, introducing attention noise and reducing representation efficiency.

This work presents a principled reformulation of generative recommendation—specifically targeting architectures like Meta’s HSTU as a ranker—by aligning sequence modeling with causal structure and attention theory. We demonstrate that the interleaving mechanism prevalent in current models acts as an inefficient proxy for similarity-weighted action pooling. To address this, we propose a structural shift that explicitly encodes the $i_n \rightarrow a_n$ causal dependency. We introduce two novel architectures, \textbf{Attention-based Late Fusion for Actions (AttnLFA)} and \textbf{Attention-based Mixed Value Pooling (AttnMVP)}, which eliminate interleaved dependencies to reduce sequence complexity by 50\%. Our framework enforces strict causal attention while preserving the expressive power of Transformer-based sequence modeling, providing a theoretically grounded and computationally efficient path for generative ranking.

\textbf{AttnLFA} performs causal attention pooling over historical actions conditioned on item similarity, whereas \textbf{AttnMVP} further integrates action signals early by mixing item and action embeddings in the Transformer value stream, progressively learning preference-aware item representations. From an information-theoretic perspective, \textbf{AttnMVP} reduces attention noise by aligning the attention space with the true causal graph of user behavior, enabling more efficient representation learning.

We evaluate our methods on large-scale product recommendation data from a major social network. Compared with the interleaved ranker baseline, AttnLFA and AttnMVP achieve consistent improvements in evaluation loss by \textbf{0.29\% and 0.8\%}, and \textbf{normalized entropy (NE)} gains across multiple tasks, while reducing training time by \textbf{23\% and 12\%} respectively. Ablation studies confirm that early, causally constrained fusion of action signals is the primary driver of performance gains. Overall, our results demonstrate that explicitly modeling item–action causality yields more efficient, scalable, and accurate generative recommender systems, offering a new design paradigm beyond token interleaving.

\end{abstract}



\received{xxxx 2026}
\received[revised]{xxxx 2026}
\received[accepted]{xxxx 2026}

\maketitle

\section{Introduction}
Generative recommender(GR) systems (e.g., Meta’s Hierarchical Sequential Transduction Units (HSTU) \cite{zhai2024actions}) model user behavior as a sequential prediction problem. These systems adopt Transformer architectures \cite{vaswani2017attention, kang2018sasrec, sun2019bert4rec, pei2021end2endbehaviorretrieval, han2025mtgr} originally developed for large language models (LLMs) \cite{kang2023llms, zhang2025recommendation, geng2022recommendation} and formulate user action prediction as a token generation process by interleaving item and action tokens. Under this formulation, user actions are generated autoregressively \cite{deng2025onerec, liang2025tbgrecall} following item tokens, enabling the model to naturally capture temporal dependencies in user interaction sequences. This paradigm represents a significant departure from traditional recommendation models \cite{naumov2019dlrm, cheng2016wide, wang2021dcnv2} and has demonstrated superior performance over conventional deep neural networks, largely due to its transductive learning capability and effective utilization of long-range user histories. As a result, generative recommendation has been widely adopted as a new modeling paradigm in industrial recommender systems.

Despite its success, the interleaving formulation introduces fundamental limitations.

\textbf{Semantic Heterogeneity of Tokens:} In natural language, tokens share a common semantic space and are compositional by nature. In contrast, recommender systems operate over fundamentally heterogeneous entities: items (e.g., posts, videos, products) and actions (e.g., click, like, share). Let $i_n \in \mathcal{I}$ denote an item token and $a_n \in \mathcal{A}$ an action token, where $\mathcal{I}$ and $\mathcal{A}$ are disjoint semantic spaces.
Interleaving these tokens into a single sequence \\
$\mathbf{x} = [i_0, a_0, i_1, a_1, \dots, i_n, a_n]$
implicitly assumes a shared latent structure across $\mathcal{I} \cup \mathcal{A}$. This assumption is weak in practice: the semantic relationship between an item and an action is asymmetric and causal, rather than compositional. Treating them as homogeneous tokens forces the Transformer to learn artificial alignments that are not grounded in the underlying data-generating process.

\textbf{Missing Explicit Causality in Self-Attention:} In our sequential framework, we model user actions through a localized causal lens. While a user's global state is informed by their interaction history $\mathcal{H}_{<n} = \{(i_k, a_k)\}_{k < n}$, we posit that the specific action $a_n$ is primarily a response to the proximal stimulus $i_n$. Formally, we define the action probability as conditioned on the current item, where the historical sequence acts as a contextual moderator rather than a direct determinant:
$$P(a_n \mid i_n, \mathcal{H}_{<n}) \approx P(a_n \mid i_n; \theta_{\mathcal{H}_{<n}})$$
Standard interleaved formulations often fail to explicitly articulate this functional mapping, treating $a_n$ and $i_n$ as part of a homogenous sequence. In the standard causal self-attention mechanism:
$$\text{Attn}(Q_n, K_{\le n}, V_{\le n})$$
where items and actions up to index $n$ contribute symmetrically to the attention computation, two primary issues arise:
\begin{itemize}
	\item \textbf{Causal Dilution:} Action tokens attend to the entire historical prefix, which dilutes the direct causal dependency on $i_n$ with unrelated historical signals.
    	\item \textbf{Structural Ambiguity:} Item tokens face difficulty in mapping specific previous actions to their corresponding items due to the uniform distribution of attention weights across all historical tokens.
\end{itemize}
Critically, positional encodings alone are insufficient to recover this structure, as they encode sequence order but lack the capacity to enforce the requisite \textbf{item-action causal pairing}.

\begin{figure}[t]
\centering
\begin{tikzpicture}[
    token/.style={draw, rectangle, minimum width=1.1cm, minimum height=0.8cm},
    attn/.style={->, thick, opacity=0.5}
]

\node[token] (i0) at (0,0) {$i_0$};
\node[token] (a0) at (1.0,0) {$a_0$};
\node[token] (i1) at (2.0,0) {$i_1$};
\node[token] (a1) at (3.0,0) {$a_1$};
\node[token] (i2) at (4.0,0) {$i_2$};
\node[token] (a2) at (5.0,0) {$a_2$};

\foreach \x in {i0,a0,i1,a1,i2} {
    \draw[attn] (a2.north) to[bend left=20] (\x.north);
}

\end{tikzpicture}
\caption{Interleaved generative recommenders treat items and actions as a single token stream. Action $a_2$ attends to all prior tokens, obscuring the direct causal dependency $i_2 \rightarrow a_2$ and introducing attention noise.}
\Description{Interleaved generative recommenders}
\label{fig:interleaved-attn}
\end{figure}
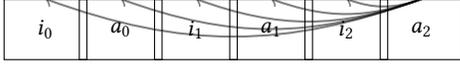

\textbf{Attention Noise Induced by Interleaving}: Even if high capacity Transformer architectures are theoretically capable of approximating latent item-action correspondences, the interleaved sequence format introduces systematic attention noise. Specifically, once the model establishes a strong causal dependency between $i_{n-1}$ and $a_{n-1}$, the subsequent token $i_n$—due to the locality-preserving nature of Rotary Position Embedding (RoPE) \cite{su2023rope} or Relative Attention Bias (RAB) \cite{raffel2023rab}—inherits a nearly identical attention bias toward $a_{n-1}$. This leads to the formation of spurious dependencies where $i_n$ attends to $a_{n-1}$ with an unwarranted inductive bias, regardless of their semantic or causal relevance. Such architectural artifacts impose an unnecessary burden on subsequent layers to 'correct' these correlations, ultimately degrading sample efficiency and complicating the optimization landscape.

\begin{figure}[t]
\centering
\begin{tikzpicture}[
    item/.style={draw, rectangle, minimum width=1.1cm, minimum height=0.8cm},
    action/.style={draw, rectangle, dashed, minimum width=1.1cm, minimum height=0.8cm},
    causal/.style={->, thick}
]

\node[item] (i0) at (0,0) {$i_0$};
\node[item] (i1) at (2.5,0) {$i_1$};
\node[item] (i2) at (5.0,0) {$i_2$};

\node[action] (a0) at (0,-1.5) {$a_0$};
\node[action] (a1) at (2.5,-1.5) {$a_1$};
\node[action] (a2) at (5.0,-1.5) {$a_2$};

\draw[causal] (i0) -- (a0);
\draw[causal] (i1) -- (a1);
\draw[causal] (i2) -- (a2);

\node at (2.5,-2.5) {\small Ground-truth causal dependency};

\end{tikzpicture}
\caption{True causal structure of user interactions. Each action $a_n$ is a response to the corresponding item $i_n$, conditioned on prior history. This structure is not explicitly represented by interleaved self-attention.}
\label{fig:true-causality}
\Description{True causal structure of user interactions}
\end{figure}

\textbf{Computational Inefficiency}: Interleaving item and action tokens increases the effective sequence length from $N$ to $2N$. Because self-attention has quadratic complexity in sequence length, this results in approximately a $4\times$ increase in both memory and computational cost. Such overhead is especially detrimental in long-horizon recommendation settings, where scalability is a primary concern. The increased sequence length substantially prolongs both training and inference, reduces GPU utilization efficiency, and leads to higher energy consumption, making interleaving-based formulations less suitable for large-scale, production-grade systems.

In this work, we propose a new formulation for Transformer-based generative recommender (GR) models that addresses the fundamental limitations of interleaved item–action sequence modeling. We first provide an explicit interpretation of interleaving-based GR methods, showing that they implicitly perform attention-based aggregation over historical user actions conditioned on item representations. Based on this observation, we reformulate GR modeling as an attention pooling mechanism in which item embeddings are used to construct the query and key projections, while action embeddings are incorporated exclusively through the value projections under a strict causal masking scheme.

Building on this formulation, we introduce two improved Transformer architectures that decouple item and action representations while preserving sequential dependency modeling. We evaluate the proposed models on large-scale product recommendation data collected from a major social networking platform and demonstrate consistent improvements over interleaving-based baselines in both recommendation accuracy and computational efficiency.
After the Transformer encoder, the item representations are discarded, and the final action token representation is extracted. This representation is concatenated with additional late-fusion features and fed into a task-specific prediction head, typically implemented using a Multi-gate Mixture-of-Experts (MMoE) architecture \cite{ma2018modeling}, to generate predictions for the target actions associated with the current item.
The item-side features comprise content and metadata representations associated with each item, including text embeddings, author embeddings, item type indicators, and related attributes. To incorporate personalization signals, we additionally include selected viewer-side contextual features, such as device type. The action features capture user feedback signals for each item, including click, skip, dwell time, like, share, and comment, etc. In our setting, user behavior is modeled using more than ten supervised action labels to represent engagement intensity and preference.
\section{Architecture overview and attention mechanism}
\label{sec:attn_mechanism}
\begin{center}
\begin{figure}
  \includegraphics[width=0.56\columnwidth]{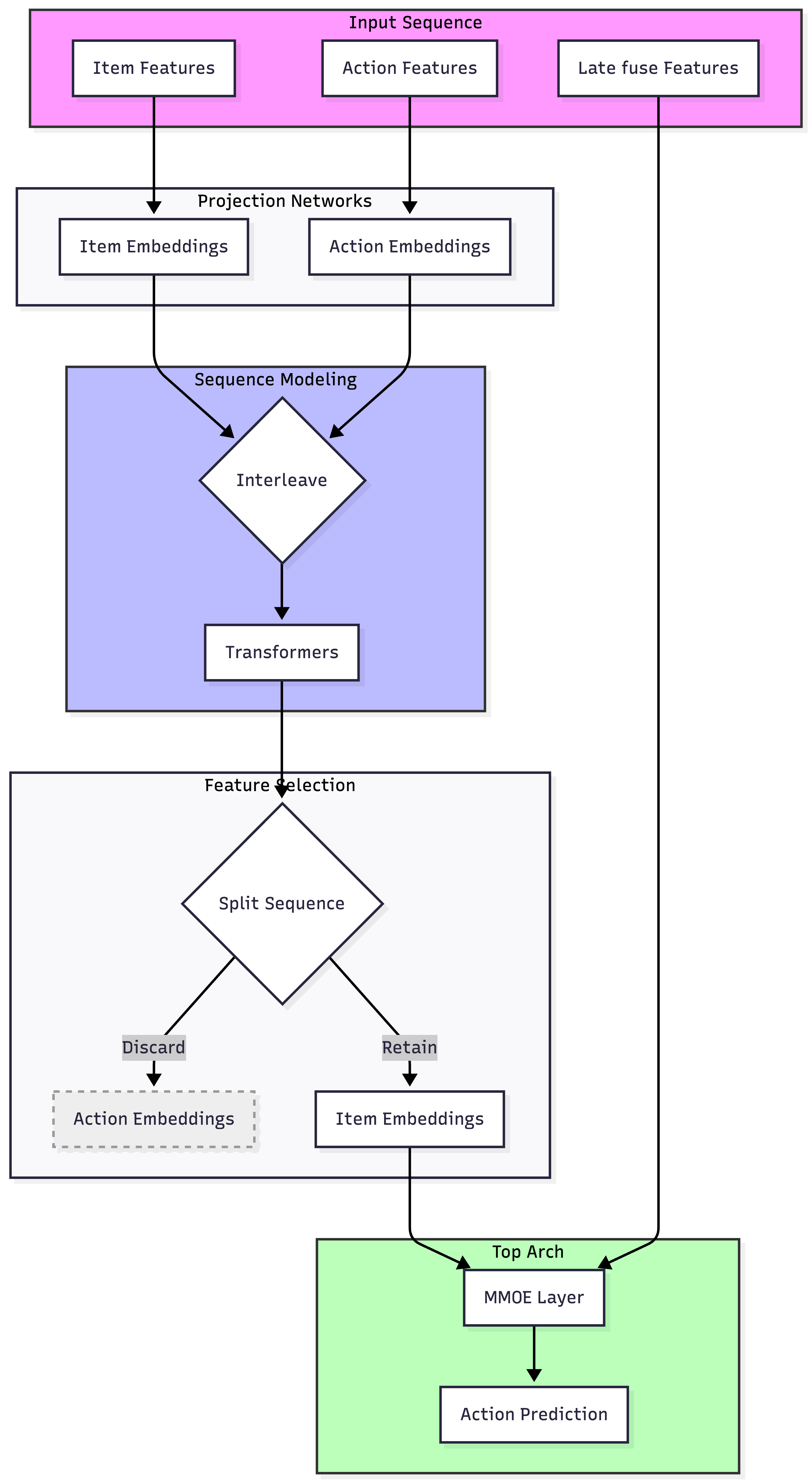}
  \caption{Traditional Generative Recommender (Interleaving Item and Action Tokens) architecture: the item and action tokens are interleaved before the transformer layers}
  \Description{Input is a sequence of Feed post and action embedding, interleaved.}
  \label{fig:gr_arch}
\end{figure}
\end{center}

Figure~\ref{fig:gr_arch} illustrates a standard Transformer-based generative recommender (GR) architecture. Raw item features are first concatenated and passed through a projection network to produce item embeddings. Similarly, action features—corresponding to multi-task supervision signals such as click, dwell time, like, share, and comment—are concatenated and projected into action embeddings using a separate projection network. The resulting item and action embeddings are then interleaved to form an input token sequence of the form $[i_0, a_0, i_1, a_1, \dots]$, which is processed by a stack of Transformer layers (12 layers in our implementation).
In addition to sequence-level representations, we employ a set of late-fusion features, primarily consisting of counting statistics. These features are introduced to improve score calibration, which is critical in production ranking systems where items are ordered according to a value function that combines multiple predicted outcomes. Stable and well-calibrated prediction scores are therefore essential for reliable and controllable ranking behavior.

For fairness and reproducibility, we use the same feature set across all model architectures evaluated in this paper. As feature engineering is not the focus of this work, we omit further details and concentrate on architectural and modeling differences in the subsequent sections.

\begin{figure}[ht]
  \centering
  \includegraphics[width=1.0\columnwidth]{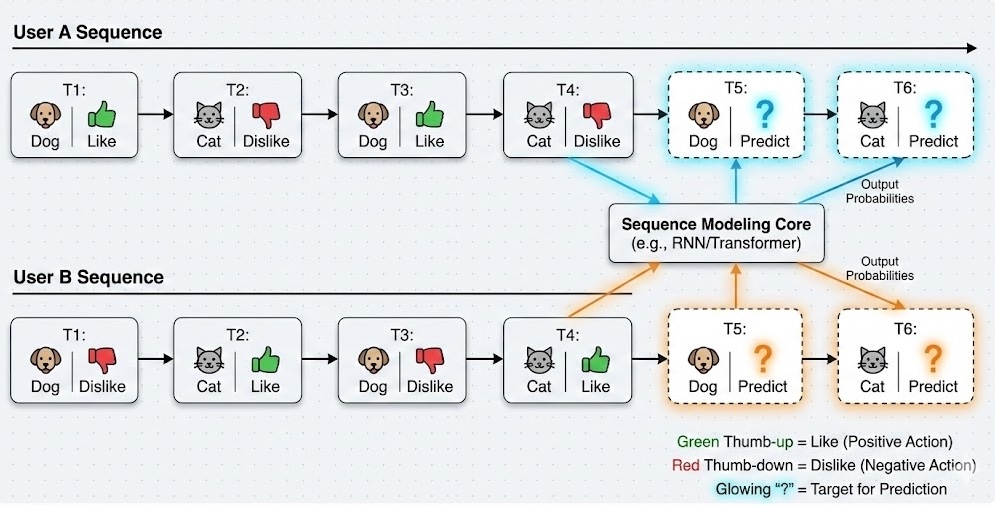}
  \caption{Illustrative toy sequences for Users A and B. The sequences demonstrate contrasting behavioral patterns: User A consistently exhibits positive interactions (e.g., "Like") with dog-related items and negative interactions with cat-related items, while User B exhibits the inverse preference profile. This highlights the model's task of capturing item-action dependencies for future state prediction.}
  \Description{User A and User B input sequence}
  \label{fig:toy_seq}
\end{figure}

We use a toy example to provide an intuitive illustration of how the generative recommender (GR) architectures operate. As shown in Figure~\ref{fig:toy_seq}, we consider two users with distinct preference profiles. User A consistently exhibits positive engagement with dog-related items and negative engagement with cat-related items, whereas User B demonstrates the opposite behavior. Given a candidate item belonging to either category, the task is to predict the user’s next action and, consequently, decide whether a dog-related or cat-related item should be presented.

In a non-sequential point-wise recommendation setting without personalization, the model would assign approximately equal probability to dog and cat items for both users, as predictions would be driven solely by aggregate item statistics across the population. To incorporate personalization in such settings, one must rely on handcrafted historical features, such as per-user like rates for dog and cat items (e.g., 100\% vs.0\% for User A, and vice versa for User B). This approach places a heavy reliance on manual feature engineering and explicit content understanding.

In particular, accurate personalization requires the model to distinguish between fine-grained item semantics (e.g., dog versus cat). If the content representation collapses both categories into a coarse concept such as “animal,” the predictive signal is substantially weakened. Moreover, achieving fine-grained semantic understanding typically necessitates extensive item taxonomies and carefully engineered historical statistics, which are costly to build and maintain at scale. These limitations motivate sequence-based generative recommender models, which aim to learn such preference patterns directly from user interaction sequences without relying on bespoke feature engineering.

Interleaved GR architectures implicitly recover such personalized signals by modeling user behavior as a single token sequence of the form $[i_0, a_0, i_1, a_1, \ldots]$. Within this formulation, the self-attention mechanism can associate an item token $i_n$ with its subsequent action token $a_n$, allowing the model to infer user preferences through repeated item–action co-occurrences. In the toy example, attention can learn that dog items are frequently followed by positive actions for User A but negative actions for User B, and vice versa for cat items. As a result, interleaving enables the model to encode user-specific preferences without explicit handcrafted features or predefined item taxonomies.

We now examine the Transformer dynamics underlying interleaved GR models in more detail. After one Transformer layer, User A’s interleaved sequence may evolve from $[i_0=\text{dog}, a_0=\text{like}]$ to contextualized representations $[dog'_0, like'_0 + \alpha \cdot dog'_0]$, where the action token aggregates information from its associated item through self-attention. From this perspective, the Transformer’s attention mechanism effectively performs a similarity-weighted pooling over historical item and action representations.

Consider a subsequent Transformer layer in which a dog-related item token $dog'_n$ attends over the historical sequence to produce its updated representation $dog''_n$. When attending to the action token $like'_0 + \alpha \cdot dog'_0$, the attention weight is amplified due to the high semantic similarity between $dog'_n$ and $dog'_0$. In contrast, action tokens associated with semantically dissimilar items (e.g., cats) receive lower attention weights. As stacking multiple Transformer layers compounds this effect, the model progressively disentangles fine-grained semantic concepts such as “liked dog,” “disliked cat,” “positive feedback on dog,” and “negative feedback on cat,” even though these abstractions are never explicitly encoded.

This analysis suggests that the effectiveness of interleaved GR models stems from using self-attention as a structured pooling operator that implicitly associates items with their corresponding user actions via semantic similarity. However, this association is formed only indirectly and incurs substantial overhead. By forcing heterogeneous item and action tokens into a single sequence, the attention mechanism must disentangle fundamentally different semantic types, introducing spurious interactions and increasing representational noise, while simultaneously doubling the effective sequence length.

For instance, in User A’s history, when a token corresponding to $cat_1$ attends to preceding tokens, it can only attend to $dog_0$ and $like_0$. As a result, its contextualized representation takes the form $cat'_1 + \beta \cdot dog'_0 + \gamma \cdot like'_0$. Even though we know User A consistently dislikes cats, this representation may still inherit positive signals associated with actions such as “like on dog” through attention, effectively encoding a “partially liked cat.” We refer to this phenomenon as attention noise, arising from indiscriminate mixing of heterogeneous tokens within interleaved self-attention.

Prior work has sought to mitigate the infrastructure and computational costs induced by interleaved generative recommender architectures. For example, Huang et al. (2025) \cite{huang2025towards} propose an early-fusion formulation in which item and action signals are embedded into a unified feature space, and a dummy action token is injected for the target item to prevent label leakage. While effective in reducing sequence length, this approach fundamentally partitions the user sequence into a context segment and a candidate segment. In the context segment, actions are treated solely as input features and cannot serve as supervision, whereas in the candidate segment, actions are used exclusively as labels and are unavailable as features. As a result, long user histories cannot be trained end-to-end; instead, the sequence must be artificially decomposed and processed in a staged or progressive manner, requiring multiple training passes and introducing additional system overhead. Moreover, the injected dummy actions in the candidate segment differ substantially from the real action embeddings used in the context segment, creating a distributional mismatch. This discrepancy distorts the attention patterns learned across segments and introduces an additional source of modeling noise, potentially degrading representation quality and convergence stability.

While Wei et al. (2025) \cite{wei2025layout} introduced Lagged Action Conditioning (LAC) to utilize $(a_{n-1}, i_n)$ pairings as input tokens, we argue that such transitions lack inherent semantic coherence. In typical recommendation environments, item sequences are exogenously determined, rendering the $(a_{n-1}, i_n)$ pairing a structural artifact rather than a valid causal transition. Despite the widespread adoption of such interleaved formats for formal consistency, the literature lacks a first-principles analysis of how Transformer architectures process these synthetic dependencies and the potential noise they introduce.

These limitations motivate a more principled formulation that preserves the expressive benefits of interleaving—namely, learning item–action associations directly from sequences—while avoiding its representational and computational drawbacks. In the next section, we introduce an alternative attention-based formulation that explicitly models this dependency without interleaving tokens.

\section{AttnLFA: Attention-based Late Fusion for Action Architecture}
\label{sec:AttnLFA}
Building on the analysis in the preceding sections, we argue that an effective generative recommender must explicitly encode the causal relationship that an exposed item $i_n$ induces a subsequent user action $a_n$. Guided by this principle, we depart from the conventional next-token prediction formulation of generative recommendation and adopt a fundamentally different perspective. Our key insight is that user actions can be modeled as a similarity-weighted aggregation over historical actions: if a target item is semantically similar to previously consumed items, then the user’s response to the target item should resemble the actions associated with those similar items.

\begin{figure}[h]
  \centering
  \includegraphics[width=0.8\columnwidth]{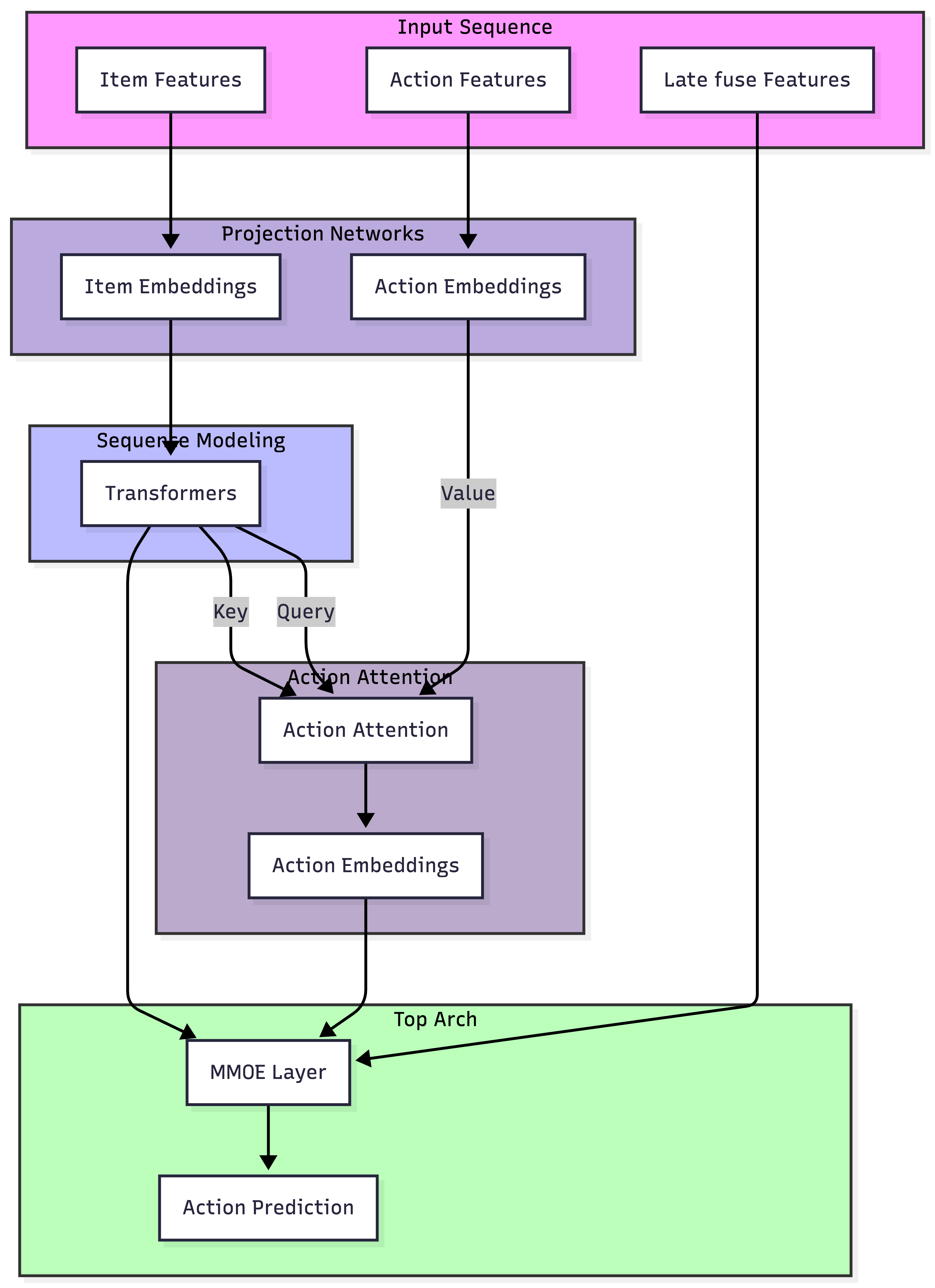}
  \caption{Attention-based Late Fusion for Action (AttnLFA). Item embeddings are transformed through a series of Transformer blocks (labeled as "Transformers" for clarity) to generate latent sequence representations. These representations serve as both Queries and Keys for the subsequent attention mechanism. In the final stage, action embeddings are integrated as Values via a causally-constrained attention pooling operation, conditioned on the sequence context. The resulting aggregated action representation is then passed to the prediction head for the final output.}
  \Description{User A and User B input sequence}
  \label{fig:late_fusion}
\end{figure}

Under this formulation, the recommendation problem is cast as an item-conditioned action pooling task, where attention serves as a structured, similarity-based pooling operator. Based on this idea, we propose the attention-based late-fusion architecture illustrated in Figure~\ref{fig:late_fusion}. Item embeddings and action embeddings are maintained as separate representation streams. Item embeddings are processed by a stack of Transformer layers to produce contextualized item representations. The final-layer item embeddings are then used as both Queries and Keys, while action embeddings are supplied as Values in the attention operation, yielding an item-conditioned pooled action representation.

To prevent label leakage, the proposed attention pooling is enforced under a strict causal constraint: the representation corresponding to item $i_n$ may attend only to items at positions $\{0, \ldots, n-1\}$, and is explicitly prohibited from attending to its own position $n$. Although such constraints can be expressed via customized attention masks, these masks are not efficiently supported by FlashAttention \cite{dao2022flashattention} kernels and significantly degrade kernel fusion and throughput.

\begin{figure}[htbp]
  \centering
  \includegraphics[width=1.0\columnwidth]{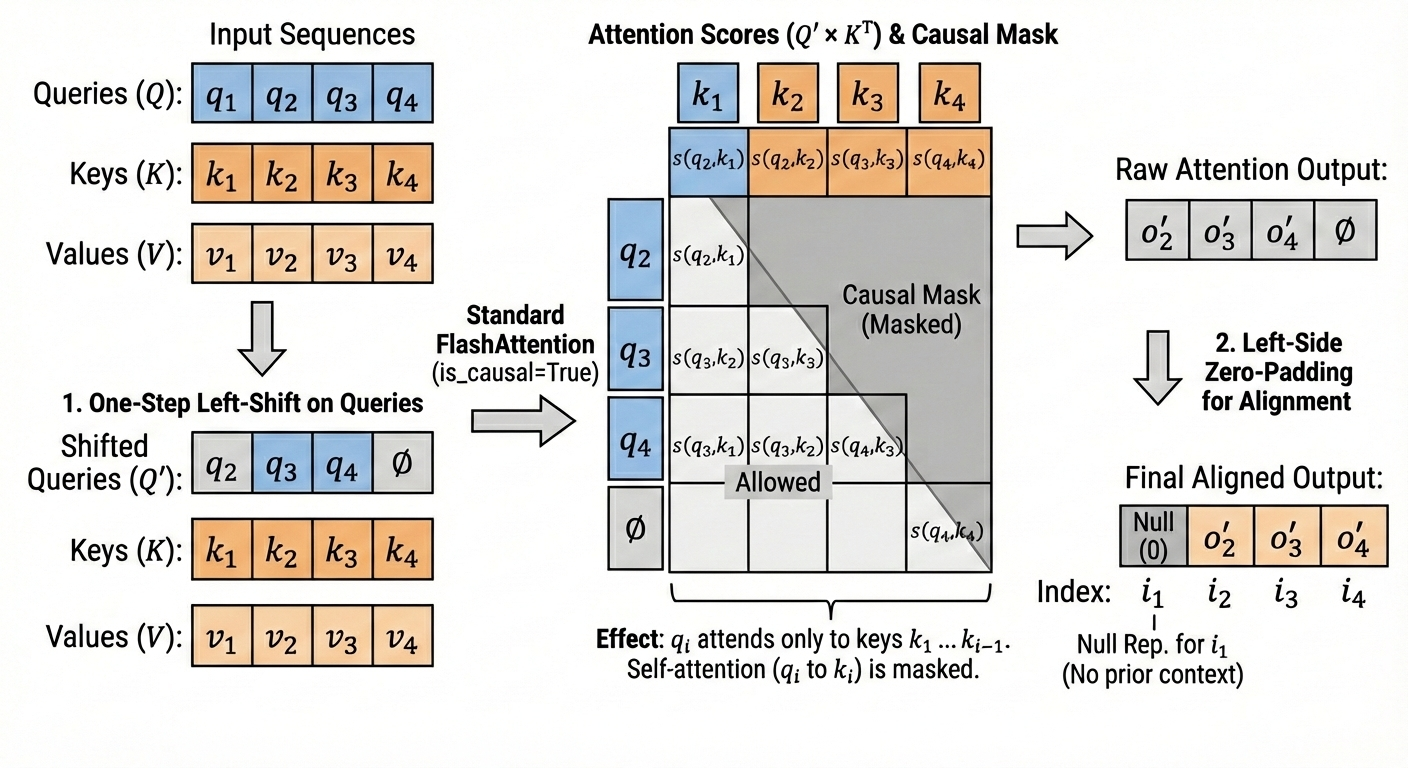}
  \caption{The \textbf{query-shifting mechanism} to enforce a strict causal constraint.}
  \label{fig:strict_causal}
\end{figure}

To leverage high-throughput GPU kernels while maintaining compatibility with standard \textbf{FlashAttention} implementations, we employ a \textbf{query-shifting mechanism} to enforce a strict causal constraint \ref{fig:strict_causal}. Specifically, we set the \verb|is_causal| flag in the module of  \verb|scaled_dot_product_attention| and apply a one-step left-shift to the query sequence $\{q_1, \dots, q_n\}$ relative to the keys. This ensures that each query $q_i$ is restricted to the preceding key prefix $\{k_1, \dots, k_{i-1}\}$, effectively preventing self-attention. Post-computation, we apply left-side zero-padding to the attention outputs to restore temporal alignment with the original sequence. Under this formulation, the first item $i_0$ naturally produces a null value representation, reflecting the absence of prior context in the sequence. This ensures structural consistency and allows for seamless integration with downstream Transformer layers, while preserving computational efficiency and numerical stability. We refer to this architecture as \textbf{Attention-based Late Fusion for Actions (AttnLFA)}.

To assess effectiveness, we compare \textbf{AttnLFA} against a strong interleaved-token baseline. Experiments are conducted on large-scale product recommendation logs collected from one of the largest professional social networks. User interaction sequences of up to 1024 events are constructed over the past 12 months and partitioned temporally. The partition immediately following the training window is reserved for evaluation. Each evaluation sequence is further divided into a \textbf{context segment}, comprising interactions occurring before the training cutoff, and a \textbf{candidate segment}, comprising interactions occurring afterward. To ensure a rigorous and controlled comparison, we maintain identical hyperparameter configurations and architectural components—including embedding layers, Transformer blocks, and projection heads—across all evaluated models, unless otherwise specified. Each model is trained for a single epoch. Furthermore, we employ \textbf{Rotary Positional Embeddings (RoPE)} \cite{su2023rope} as the universal positional encoding scheme throughout our experiments to ensure consistency in spatial modeling.

To faithfully approximate online serving conditions, we apply a \textbf{timestamp-based label masking scheme} during evaluation. Loss and metrics are computed exclusively on candidate items. For the context segment, \textbf{standard causal masking} is applied to interleaved baseline, while \textbf{strict causal masking} is enforced for \textbf{(AttnLFA)} . For the candidate segment, candidate items are prohibited from attending to one another. This evaluation protocol prevents information leakage while maintaining consistency with real-world recommendation constraints.

\begin{table}[htbp]
\centering
\begin{tabular}{@{}llllll@{}}
\toprule
Model & \shortstack{Eval \\ Loss} & \shortstack{LongDwell \\ NE} & \shortstack{Contribution \\ NE} & \shortstack{Like \\NE} & Time\\ \midrule
Baseline &  -   &  - & -  & - & - \\
AttnLFA & \textbf{-0.29\%} & \textbf{-0.06\%} & \textbf{-0.49\%}  & \textbf{-0.47\%} & \textbf{-22.8\%}\\
\end{tabular}
\caption{Performance comparison between AttnLFA and Baseline. We report the relative improvement across four key dimensions: (a) Multi-task Binary Cross Entropy (BCE) Loss, (b) evaluation Normalized Entropy (NE) for Long Dwell, Contribution, and Like actions, and (c) total training latency. AttnLFA demonstrates superior predictive accuracy (lower NE/Loss) while maintaining competitive computational efficiency.}
\label{tab:lfa_result}
\vspace{-2.5em}
\end{table}

We evaluate model performance across three primary engagement signals, formulated as binary classification tasks:
\begin{itemize}
\item Long Dwell: A binary indicator $\mathbf{1}(dwell\_time > \tau)$, where $dwell\_time$ is the user's dwell time and $\tau$ is a predefined temporal threshold.
\item Contribution: A multi-action signal where the label is positive if the user performs at least one non-click engagement (e.g., like, comment, or share).
\item Like: A specific binary task representing whether the user explicitly engaged with the "like" mechanism.
\end{itemize}
Table \ref{tab:lfa_result} summarizes the experimental results, where all multi-task models use the binary labels in each task, and are optimized using standard \textbf{Binary Cross Entropy (BCE) loss}. For brevity, we report performance on three representative major tasks; the remaining nine tasks exhibit consistent performance trends and are omitted for conciseness. \textbf{AttnLFA} achieves substantial improvements in  \textbf{evaluation loss} and \textbf{normalized entropy (NE)} across the primary prediction tasks. In addition, by eliminating the interleaving formulation, the proposed approach reduces end-to-end training time by \textbf{22.8\%}, demonstrating both modeling and computational efficiency gains. The observed improvements corroborate our central hypothesis: accurately modeling the causal relationship from item $i_n$ to action $a_n$ is critical for improving predictive performance. In \textbf{AttnLFA}, this causality is explicitly enforced by design—each action representation $a_n$ is obtained by attention-based pooling driven solely by the corresponding item representation $i_n$. This mechanism operationalizes a simple but powerful principle: items that are semantically similar should induce similar action distributions.

Encouraged by the effectiveness of \textbf{AttnLFA}, we further extend this idea to a more expressive variant. Specifically, instead of applying action embeddings through late fusion, we explore an early-fusion formulation that integrates item–action interactions at earlier stages of representation learning.

\section{AttnMVP: Attention based Mixed Value Pooling Architecture}
\label{sec:AttnMVP}

\begin{figure}[htbp]
  \centering
  \includegraphics[width=0.8\columnwidth]{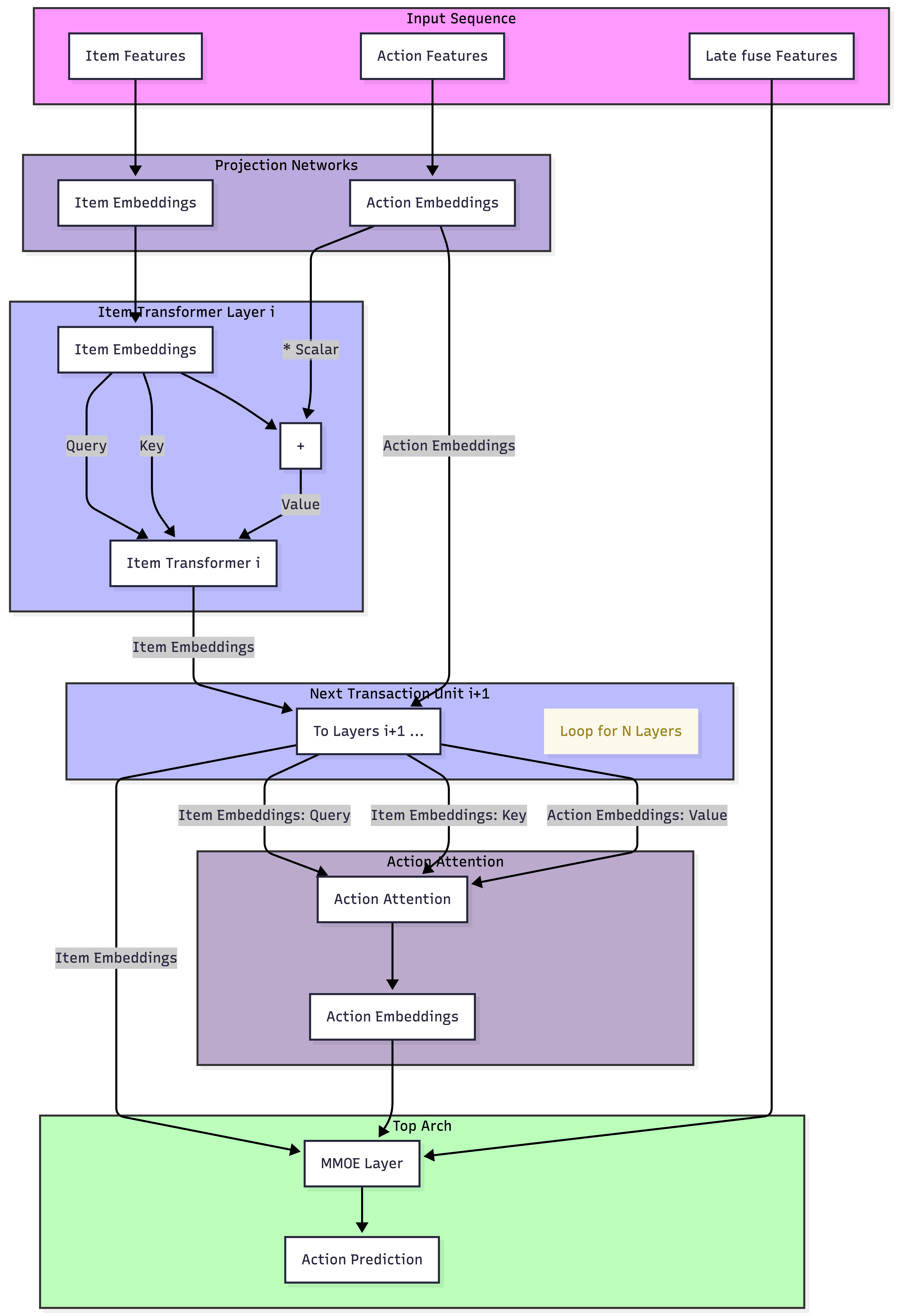}
  \caption{Attention-based Mixed Value Pooling (AttnMVP) architecture. Item embeddings serve as Queries and Keys in each Transformer layer, while item and action embeddings are additively fused as mixed Values. Across stacked layers, action signals are progressively injected into item representations under strict causal constraints. In the final stage, action embeddings are pooled via causally masked attention conditioned on the sequence-level item representations, and the pooled action representation is fused with the final item embedding to produce action predictions.}
  \Description{Attention-based Mixed Value Pooling}
  \label{fig:attnMVP}
\end{figure}

Building on the effectiveness of AttnLFA, we introduce an early-fusion variant termed Attention-based Mixed Value Pooling \textbf{(AttnMVP)}, shown in Figure \ref{fig:attnMVP}. \textbf{AttnMVP} reformulates action-conditioned sequential recommendation as a causally constrained representation learning problem, in which item representations are iteratively refined by integrating historical user actions through attention-based value mixing.

Let $\{\mathbf{i}_t\}_{t=1}^T$ denote the sequence of item embeddings and $\{\mathbf{a}_t\}_{t=1}^T$ the corresponding action embeddings (e.g., click, dwell, like), following the conventions of SASRec. At Transformer layer $\ell$, AttnMVP applies self-attention over item representations using
$\mathbf{Q}^{(\ell)} = \mathbf{K}^{(\ell)} = \mathbf{H}^{(\ell-1)}$,
where $\mathbf{H}^{(0)} = \{\mathbf{i}_t\}$ and $\mathbf{H}^{(\ell)}$ denotes the item representations after layer $\ell$. The value vectors are constructed via mixed-value fusion:
$\mathbf{V}^{(\ell)}_t = \mathbf{H}^{(\ell-1)}_t + \lambda \mathbf{a}_t$,
where $\lambda \ge 0$ controls the contribution of action signals. We adopt an additive fusion for $\mathbf{H}^{(\ell-1)}_t$ and $\lambda \mathbf{a}_t$ primarily to prioritize computational efficiency and maintain a lightweight architecture. While this linear combination minimizes overhead, exploring more sophisticated fusion mechanisms—such as gated pooling —remains a promising avenue for future research and is expected to yield further performance gains. In practice, we set $\lambda = 1$, and preliminary sensitivity analysis over $\lambda \in [0.5, 1.0]$ indicates stable performance. After $T$ (= 12) layers of Transformer blocks, at the final layer, we apply an action pooling operation identical to \textbf{AttnLFA}. This formulation enables each Transformer layer to perform causally masked, attention-weighted aggregation of historical action signals into item representations, conditioned on item similarity. As item embeddings propagate through successive layers, they evolve from encoding generic content semantics (e.g., dog versus cat) to capturing user-conditioned semantics (e.g., preferred dog versus disfavored cat). Through this progressive integration, user preference signals are implicitly embedded into the item representations, amplifying item–item contrast in a personalized manner. Notably, this personalization emerges end-to-end from the attention mechanism itself, without requiring explicit user profiling or handcrafted personalization features.

From a representation learning perspective,  \textbf{AttnMVP} explicitly encodes the inductive bias that semantically related items elicit analogous user responses. By decoupling item and action representations, the model circumvents the heterogeneous token entanglement and quadratic computational overhead inherent in interleaved generative frameworks. Consequently, AttnMVP offers a principled and scalable alternative to token-level interleaving for sequential behavioral modeling.

\begin{table}[htbp]
\centering
\begin{tabular}{@{}lclclclclcl@{}}
\toprule
Model & \shortstack{Eval \\ Loss} & \shortstack{LongDwell \\ NE} & \shortstack{Contribution \\ NE} & \shortstack{Like \\NE} & Time\\ \midrule
Baseline &-&-& -  & - & - \\
AttnMVP & \textbf{-0.80\%} & \textbf{-0.41\%} & \textbf{-1.1\%}  & \textbf{-1.1\%} & \textbf{-12.3\%}\\
 \shortstack{AttnMVP \\ - LFA} & \textbf{-0.78\%} & \textbf{-0.40\%} & \textbf{-1.0\%}  & \textbf{-1.0\%} & \textbf{-13.02\%}\\
\end{tabular}
\caption{Relative performance improvement of Eval Loss, major tasks' Eval Normalized Entropies (NEs) and Training Time for Baseline, AttnMVP and AttnMVP without LFA. (AttnMVP-LFA)}
\label{tab:mvp_result}
\vspace{-2.5em}
\end{table}

Table~\ref{tab:mvp_result} reports the relative improvements in \textbf{evaluation loss}, \textbf{normalized entropy (NE)} for major prediction tasks, and \textbf{training time}. Compared with the interleaved ranker baseline, \textbf{AttnMVP} delivers consistent and larger gains in both loss and NE across all tasks, while also reducing training time by \textbf{12.3\%}. These improvements exceed those achieved by \textbf{AttnLFA}, indicating the significant benefit of integrating action information earlier in the representation learning process.

To isolate the contribution of early fusion, we further evaluate a variant of \textbf{AttnMVP} that retains mixed-value fusion within Transformer layers but removes the late fusion attention (denoted as \textbf{AttnMVP–LFA}). This variant achieves performance comparable to the full model, with only marginal degradation in loss and NE. The result suggests that the majority of the gains stem from early, causally constrained integration of action signals into item representations. This finding supports our hypothesis that explicitly encoding preference-aware semantics—e.g., distinguishing “preferred” versus “disfavored” items—within the sequence representations is critical for effective generative recommendation modeling.

\section{Future Work: AttnDHN - Attention based Dual-Helix Network}
\label{sec:AttnDHN}
\begin{figure}[htbp]
  \centering
  \includegraphics[width=0.8\columnwidth]{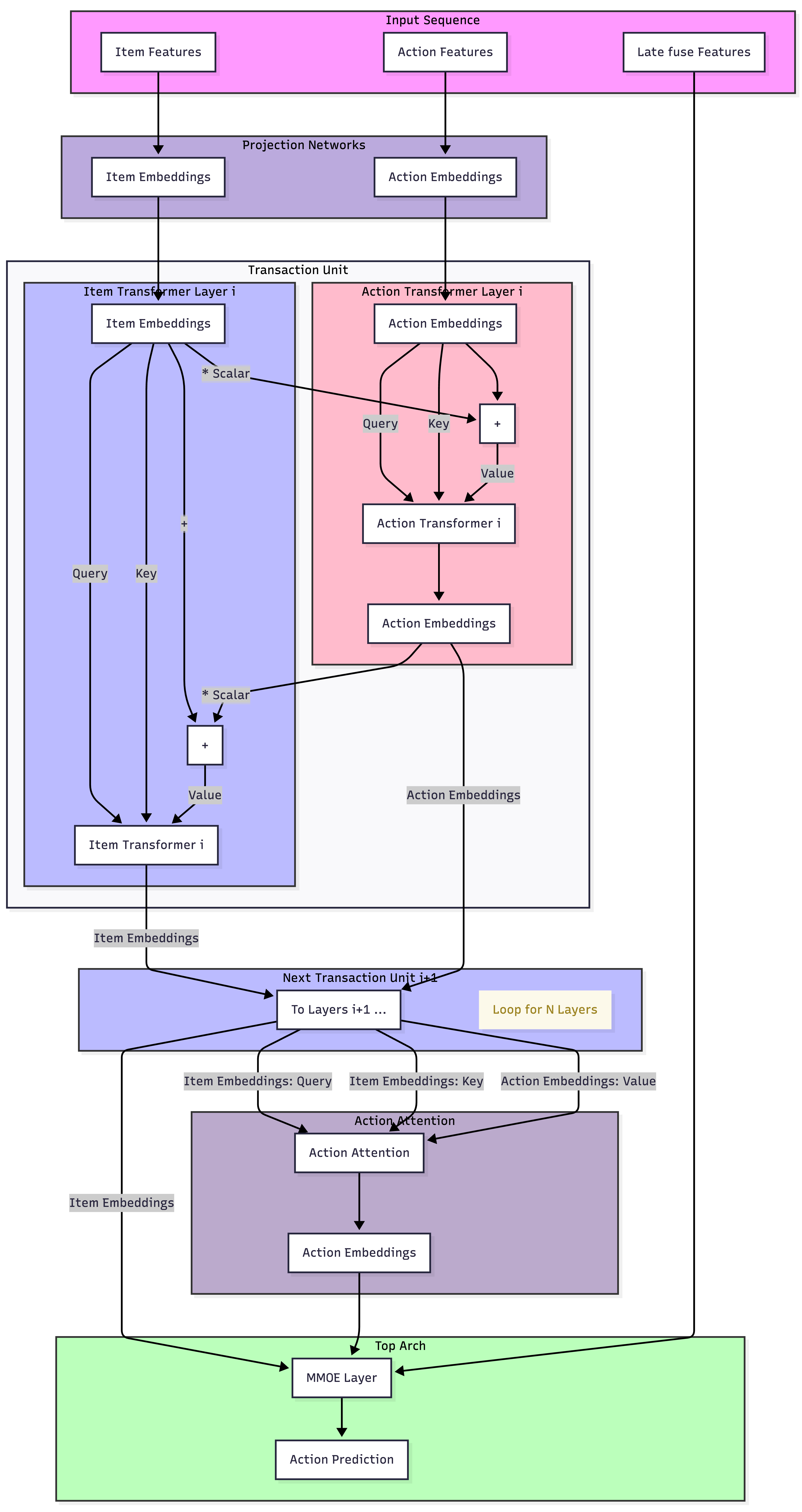}
  \caption{Attention-based Dual-Helix Network (AttnDHN) architecture. Action embedding and Item embeddings are updated in pair-wise sequence in individual Transformer layer. Both transformer layer use either action and item embedding as query and key, and use a combination of item + action embedding as value. In the final stage, action embeddings are pooled via causally masked attention conditioned on the sequence-level item representations, and the pooled action representation is fused with the final item embedding to produce action predictions.}
  \Description{Attention-based Mixed Value Pooling}
  \label{fig:attnDNA}
\end{figure}

Motivated by the strong empirical performance of \textbf{AttnMVP}, we further propose a symmetric dual-stream architecture, termed Attention-based Dual-Helix Network \textbf{(AttnDHN)}. In \textbf{AttnMVP}, item representations are updated via self-attention with mixed item–action values, using $(Q_t, K_t, V_t) = (i_t, i_t, i_t + a_t)$. \textbf{AttnDHN} extends this formulation by introducing a complementary action-centric update, in which action representations are simultaneously refined using $(Q_t, K_t, V_t) = (a_t, a_t, i_t + a_t)$. The details can be referred from Figure \ref{fig:attnDNA}.  Within each Transformer block, item and action streams are updated sequentially in a paired manner, forming a tightly coupled interaction unit. This alternating update mechanism induces a bidirectional flow of information between item and action representations, analogous to a double-helix structure, enabling more expressive and coherent co-evolution of user preference signals across network depth.

To date, we do not observe \textbf{AttnDHN} to consistently outperform \textbf{AttnMVP}. We attribute this outcome to three primary factors. First, \textbf{AttnDHN} exhibits reduced training stability relative to \textbf{AttnMVP}; in practice, stable optimization requires halving the learning rate. When both models are trained for the same number of optimization steps, this constraint leads to weaker convergence and inferior performance for \textbf{AttnDHN}. Second, \textbf{AttnDHN} effectively doubles the number of Transformer updates per layer due to its dual-stream design, making direct comparisons with \textbf{AttnMVP} less straightforward. Improvements observed at a fixed depth (e.g., 12 layers) cannot be cleanly attributed to architectural superiority, as similar gains might be achievable by increasing the depth of \textbf{AttnMVP}. Third, and more fundamentally, item and action tokens reside in highly heterogeneous semantic spaces: the action vocabulary is small (on the order of tens), whereas the item space is effectively unbounded. As a result, action-centric representations tend to be less expressive and noisier than item-centric representations (e.g., “preferred dog” versus a diffuse mixture of preferences over many unrelated items).

Despite these limitations, we include \textbf{AttnDHN} as an exploratory architecture. Its symmetric dual-stream design may be better suited to settings in which the two representation spaces are more homogeneous, such as multimodal recommendation scenarios that jointly model text and visual embeddings. We leave a systematic investigation of such applications to future work.

\section{Conclusion}
\label{sec:conclusion}
We revisit the interleaved-token formulation common in generative recommendation and offer a first-principles critique of its operational mechanics. Our analysis reveals that while self-attention effectively acts as a latent pooling mechanism for user actions via item-level semantics, the standard interleaved approach remains suboptimal. Specifically, this formulation introduces representational noise and computational inefficiency by interleaving heterogeneous tokens, which not only doubles sequence length but also complicates the attention landscape.

Guided by the causal structure that an item interaction $i_t$ induces its corresponding user action $a_t$, we propose a family of attention-based architectures that explicitly encode this causality without interleaving. We introduce \textbf{AttnLFA}, which formulates action modeling as causally masked attention-based late fusion, and \textbf{AttnMVP}, which further generalizes this idea through mixed-value early fusion that progressively integrates historical action information into item representations. From an information-theoretic perspective, these designs reduce attention noise by constraining aggregation to causally valid and semantically aligned interactions, enabling more efficient and expressive representation learning.

Empirically, both architectures consistently outperform a strong interleaved ranker baseline on large-scale real-world recommendation data, achieving lower \textbf{evaluation loss} and \textbf{normalized entropy} across major tasks while substantially reducing training time and computational cost. These gains validate our central thesis: explicitly modeling the causal relationship from items to actions leads to both improved predictive accuracy and better system efficiency.

Finally, we explore a symmetric dual-stream extension, \textbf{AttnDNA}, and discuss its limitations in standard recommender settings due to semantic heterogeneity between item and action spaces, while highlighting its potential applicability to more homogeneous multimodal scenarios. Overall, our results suggest that moving beyond interleaving toward causality-aware attention formulations offers a principled and scalable path forward for generative recommender systems.

\begin{acks}
The author gratefully acknowledges Samaneh Moghaddam and Ying Xuan for funding and support. Special thanks to Chen Zhu, Tao Huang, and Antonio Alonso for their guidance and discussions on model training. The author also thanks the LinkedIn Feed AI team, Apollo Engineering team and Core AI team for valuable discussions and infrastructure support.
\end{acks}

\bibliographystyle{ACM-Reference-Format}
\bibliography{base}

\end{document}